\documentclass[prd,aps,showpacs,tightenlines, 10pt, reprint,nofootinbib,superscriptaddress,floatfix,twocolumn]{revtex4-1}

\usepackage{verbatim}
\usepackage{natbib}
\usepackage{graphicx}
\usepackage{amsmath,amssymb}
\usepackage{bm}
\usepackage{dsfont}
\usepackage{cancel}
\usepackage{float}
\usepackage{siunitx}
\usepackage{xspace}
\usepackage{comment}
\usepackage{makecell}
\usepackage{xcolor}
\usepackage{dcolumn}
\usepackage{multirow}

\usepackage[colorlinks=true,citecolor=blue,urlcolor=blue,linktocpage=true,linkcolor=blue]{hyperref}
\usepackage[capitalize]{cleveref}
\usepackage[absolute,overlay]{textpos}

\usepackage[utf8]{inputenc} 
\usepackage{acronym}
\usepackage{amsthm}  
\usepackage{mathtools}
\usepackage[section]{placeins}
\usepackage{orcidlink}
\usepackage{physics} 
\usepackage[normalem]{ulem}  
\usepackage{array}
\newcolumntype{C}[1]{>{\centering\let\newline\\\arraybackslash\hspace{0pt}}m{#1}}
\usepackage{cellspace}
\usepackage[export]{adjustbox} 
\usepackage{csquotes}
\usepackage{slashed}
\usepackage{bbold}
\hypersetup{
    colorlinks,
    linkcolor={red!60!black},
    citecolor={blue!50!black},
    urlcolor={blue!80!black}
}

\newcommand*{\vect}[1]{\ensuremath{\mathbf{#1}}\xspace}
\newcommand{\rmi}[1]{{\mbox{\scriptsize #1}}}
\newcommand{\rmii}[1]{{\mbox{\tiny\rm{#1}}}}
\newcommand{\MPl}{M_\rmii{Pl}}

\newcommand{\GeV}{\mathrm{GeV}}

\newcommand{\rsc}{r_\rmi{sc}}

\newcommand{\fphi}{f_\phi}
\newcommand{\mphi}{m_\phi}

\newcommand{\Tosc}{T_\rmi{osc}}
\newcommand{\Hosc}{H_\rmi{osc}}

\newcommand{\Toscmis}{T_\rmi{osc}^\rmi{mis}}
\newcommand{\Hoscmis}{H_\rmi{osc}^\rmi{mis}}

\newcommand{\rhophiosc}{\rho_{\phi,\mathrm{osc}}}
\newcommand{\rhophistar}{\rho_{\phi,\star}}

\newcommand{\mD}{m_\rmii{D}}
\newcommand{\astar}{a_\star}

\newcommand{\alphamin}{\alpha_\rmi{min}}
\newcommand{\Trh}{T_\rmi{rh}}

\newcommand{\TBBN}{T_\rmii{BBN}}

\newcommand{\led}{\lambda_{\mathrm{ed}}}

\newcommand{\arec}{a_{\rmi{rec}}}
\newcommand{\aed}{a_{\rmi{ed}}}

\begin{document}
\begin{textblock*}{\textwidth}[1,0](200mm,10mm)
\noindent
\flushright
KA-TP-10-2026\\
MITP-26-025
\end{textblock*}

\title{Audible Axion Magnetogenesis: \\Linking Intergalactic Magnetic Fields and Gravitational Waves}

\author{Christopher Gerlach}
\email{cgerlach@uni-mainz.de} 
\affiliation{PRISMA$^{++}$ Cluster of Excellence \& Mainz Institute for Theoretical Physics, Johannes Gutenberg-Universität Mainz, 55099 Mainz, Germany}

\author{Daniel Schmitt}
\email{daniel.schmitt@kit.edu} 
\affiliation{Institute for Astroparticle Physics (IAP), Karlsruhe Institute of Technology (KIT), Hermann-von-Helmholtz-Platz 1, 76344 Eggenstein-Leopoldshafen, Germany}
\affiliation{Institute for Theoretical Physics (ITP), Karlsruhe Institute of Technology (KIT), Wolfgang-Gaede-Str.~1, 76131 Karlsruhe, Germany}
\affiliation{Institute for Theoretical Physics, Goethe University, 60438 Frankfurt am Main, Germany}

\author{Pedro Schwaller}
\email{pedro.schwaller@uni-mainz.de} 
\affiliation{PRISMA$^{++}$ Cluster of Excellence \& Mainz Institute for Theoretical Physics, Johannes Gutenberg-Universität Mainz, 55099 Mainz, Germany}

\date{\today}

\begin{abstract}
Identifying dark matter candidates that simultaneously generate multiple observable cosmological signatures is a key goal in connecting particle physics with upcoming observations. Axion-like particles coupled to the Standard Model photon offer a promising framework. 
In the trapped misalignment mechanism, the onset of axion oscillations is delayed, inducing a period of supercooling in the early Universe.
This can lead to exponential production of photon quanta via a tachyonic instability, generating observable gravitational wave signatures. 
Simultaneously, reheating of the Standard Model plasma produces strong, helical magnetic fields on intergalactic scales. 
The parameter space most promising for gravitational wave detection yields magnetic field strengths that exceed lower bounds from blazar observations.
\end{abstract}

\maketitle

\section{Introduction}
Axions~\cite{Peccei:1977hh,Peccei:1977ur,Weinberg:1977ma,Wilczek:1977pj} and axion-like particles (ALPs) are among the most compelling dark matter candidates~\cite{Preskill:1982cy,Abbott:1982af,Dine:1982ah}, may provide a framework for realizing cosmic inflation~\cite{Freese:1990rb,Pajer:2013fsa,Adshead:2015pva,Daido:2017wwb,Takahashi:2021tff,Kitajima:2021bjq}, and naturally arise in many ultraviolet completions of the Standard Model~(SM)~\cite{Witten:1984dg,Svrcek:2006yi,Arvanitaki:2009fg,Marsh:2015xka}.
Unfortunately, the regime of large decay constants $\fphi$, where the ALP becomes effectively invisible, remains challenging to probe experimentally~\cite{OHare:2024nmr}.
In the audible axion model, an oscillating ALP triggers a tachyonic resonance in the equation of motion of an Abelian gauge boson, leading to copious production of gauge quanta in the early Universe~\cite{Agrawal:2017eqm,Machado:2018nqk,Machado:2019xuc,Co:2019jts,Chang:2019tvx,Co:2021rhi,Chatrchyan:2020pzh,Salehian:2020dsf,Namba:2020kij,Kite:2020uix,Kitajima:2020rpm,Ratzinger:2020oct,Banerjee:2021oeu,Madge:2021abk,Eroncel:2022vjg,Madge:2023dxc,Su:2025nkl,Hook:2016mqo,Gerlach:2025fkr}; see also~\cite{Anber:2009ua,Anber:2012du,Barnaby:2012xt,Domcke:2016bkh,
Greene:1997fu,Kofman:1994rk,Shtanov:1994ce,Kofman:1997yn,Dufaux:2007pt,Maleknejad:2016qjz,Figueroa:2016wxr,Adshead:2018doq,Cuissa:2018oiw,Ramberg:2022irf,Cui:2023fbg,Cyncynates:2023zwj,Cyncynates:2024yxm,Xu:2024kwy,Ramberg:2025nxt}.
This induces large anisotropies in the cosmic fluid, resulting in the production of a stochastic gravitational wave~(GW) background and thereby opening the large-$f_\phi$ parameter space to future GW observatories.

We have shown that in the case of ALP trapped misalignment~\cite{Higaki:2016yqk,Kawasaki:2017xwt,Nakagawa:2020zjr,DiLuzio:2021pxd,DiLuzio:2021gos,Kitajima:2023pby,DiLuzio:2024fyt}, where the onset of ALP oscillations is delayed, a coupling between an ALP and the SM photon suffices to source an efficient tachyonic resonance~\cite{Gerlach:2025fkr}.
Taking into account cosmological constraints and the impact of Schwinger pair production~\cite{Heisenberg:1936nmg,Schwinger:1951nm,Kobayashi:2014zza,Hayashinaka:2016qqn,Gould:2017fve,Lozanov:2018kpk,Domcke:2018eki,Domcke:2019qmm,Domcke:2021yuz}, we have identified two distinct regimes at small and very large ALP masses $\mphi$ where significant GW production takes place.
In this work, we show that in the small-$\mphi$ regime, large-scale helical magnetic fields produced through the tachyonic resonance may survive until today, constituting intergalactic magnetic fields.
Previously, ALP-induced magnetogenesis was studied during inflation~\cite{Fujita:2015iga,Adshead:2016iae,Patel:2019isj,Gorbar:2021zlr,Talebian:2021dfq,Anzuini:2024rpl,Brandenburg:2024awd,Sharma:2024nfu} and after recombination~\cite{Brandenberger:2025gks,Kamali:2026tgq} where plasma effects can be neglected.
We show that trapped ALPs can sufficiently suppress the temperature of the plasma, opening up a large parameter space for primordial magnetogenesis.

Remarkably, the parameter space that yields the strongest GW signals also generates magnetic fields large enough to account for the intergalactic magnetic fields implied by recent blazar observations~\cite{Ando:2010rb,Tavecchio:2010mk,doi:10.1126/science.1184192,Taylor:2011bn,Essey:2010nd,Chen:2014rsa,Fermi-LAT:2018jdy}. 
This establishes a direct link between future GW measurements and existing astrophysical data, rendering the audible axion model testable via multi-messenger techniques.

\section{Supercooled audible axions}\label{sec:setup}
In the following, we briefly introduce our setup. We study an ALP $\phi$ coupled to the SM photon $A_\mu$, hence the action reads
\begin{align}
    {\cal S}  = \int &d^4x \sqrt{-g}\left[ \frac{1}{2}\partial_\mu \phi \partial^\mu \phi - V(\phi) \right. \notag\\ 
     &\left. - \frac{1}{4}F_{\mu\nu}F^{\mu\nu} - \frac{\alpha}{4 f_\phi} \phi F_{\mu\nu}\tilde{F}^{\mu\nu} \right] \, .
\end{align}
Here, $V(\phi)$ is the ALP potential, $F_{\mu\nu}$~($\tilde{F}_{\mu\nu}$) denotes the electromagnetic (dual) field strength tensor, and $g=\det(g_{\mu\nu})$ with the Friedmann-Lemaître-Robertson-Walker (FLRW) metric~$ds^2 = g_{\mu\nu} dx^{\mu}dx^\nu = a(\tau)^2 (d \tau^2 - \delta_{ij} x^ix^j)$, where $a$ is the scale factor as function of conformal time $\tau$.
The coupling strength between the ALP and the SM photon is parametrized by $\alpha$, and $\fphi$ is the ALP decay constant.
We assume the pre-inflationary scenario, i.e., the ALP is homogeneous.

We follow~\cite{Gerlach:2025fkr} and consider an ALP potential of the form
\begin{equation}
    V(\phi) = \mphi^2 \fphi^2 \left(1-\cos\left(\frac{\phi}{\fphi}\right)\right) + V_{\cancel{\rmii{PQ}}} \, ,
\end{equation}
where $\mphi$ is the ALP mass.
Apart from to the usual cosine contribution to the potential, we introduce additional explicit $U(1)_\rmii{PQ}$-breaking operators via $V_{\cancel{\rmii{PQ}}}$.\footnote{We remain agnostic about the exact shape of $V_{\cancel{\rmii{PQ}}}$. For explicit realizations, see \cite{DiLuzio:2024fyt}.}
Such operators can induce additional minima in the ALP potential separated by (thermal) barriers, which trap the pseudoscalar initially~\cite{Higaki:2016yqk,Kawasaki:2017xwt,Nakagawa:2020zjr,DiLuzio:2021pxd,DiLuzio:2021gos,Kitajima:2023pby,DiLuzio:2024fyt,Ramberg:2025nxt}.
Then the onset of ALP oscillations is delayed with respect to the ordinary misalignment mechanism, which is in the following referred to by the superscript \enquote{mis}.
This delay can be parametrized by the supercooling ratio
\begin{equation}
    \rsc = \frac{\Tosc}{T_\rmi{osc}^\rmi{mis}} \, ,
\end{equation}
where $T_\rmi{osc}^\rmi{mis}$ is determined by equating the Hubble parameter and the ALP mass, $\Hoscmis = \mphi$.
By adopting a specific trapping potential, we have shown that $\rsc~\ll~1$ can easily be achieved~\cite{Gerlach:2025fkr}.
In the following, we therefore assume that the ALP is trapped in some false minimum at $\phi_i = \theta \fphi$ with $\theta \sim 1$, until oscillations start at the temperature $\Tosc \ll \Toscmis$, which is fixed below.

To show the effect of trapped misalignment on the dynamics of the ALP-photon system, let us consider the equation of motion of the photon. 
We decompose the gauge field into Fourier modes $v_\pm(k,\tau)$, where $\pm$ corresponds to the respective helicity.
We obtain
\begin{align}\label{eq:photon_eom}
    v_\pm ''(k,\tau) + \omega^2_\pm (k,\tau) v_\pm(k,\tau) & = 0\,,
\end{align}
with $'$ denoting derivatives with respect to $\tau$. 
The dispersion relation of the photon, which is in thermal equilibrium with the primordial plasma, reads~\cite{Kraemmer:1994az,Kapusta:2006pm,Bellac:2011kqa}
\begin{align}\label{eq:dispersion_full}
    \omega^2 - k^2 \mp k\frac{\alpha}{f_\phi}\phi'
    = \frac{a^2\mD^2}{2}\left[\frac{\omega}{2k}\ln\left(\frac{\omega+k}{\omega-k}\right)\right.
    \\ \nonumber
    \left. -\frac{\omega^3}{2k^3}\ln\left(\frac{\omega+k}{\omega-k}\right)+
    \frac{\omega^2}{k^2}
    \right]\,,
\end{align}
where $\pm$ indicates the respective helicity.\footnote{Note that this expression assumes an $e^+e^-$ plasma. This is justified, since we will only consider scenarios where photon production occurs at low temperatures~(see below), such that muons are strongly Boltzmann-suppressed.}
Here, $\mD = eT/\sqrt{3}$ is the Debye mass with electromagnetic charge $e=0.3$.
Notably, the onset of ALP oscillations, $\phi'~\neq~0$, modifies the photon dispersion relation. 
A tachyonic band emerges; modes within the range $0 < k < \alpha\phi'/\fphi$ become unstable and grow exponentially, $v \propto \exp(\omega_\rmii{T} \tau)$, with the comoving peak growth rate given by~\cite{Gerlach:2025fkr}
\begin{equation} \label{eq:peak_growth_rate_SM}
    \tilde{\omega}_\rmii{T}  \approx \frac{16}{9 \pi} \frac{(\alpha\theta\mphi)^3}{(e T)^2} a \,.
\end{equation}
This holds for one helicity at a time, depending on the sign of $\phi'$.

Note that in the ordinary misalignment mechanism, this growth rate is typically orders of magnitude smaller than the Hubble rate,
\begin{equation}
    \frac{\tilde{\omega}_\rmii{T}(\Toscmis)}{a \Hoscmis} \sim \left(\frac{\mphi}{\Toscmis}\right)^2 \sim \left(\frac{\Hoscmis}{\Toscmis}\right)^2 \sim \left(\frac{\Toscmis}{\MPl}\right)^2 \, ,
\end{equation}
preventing efficient tachyonic growth. 
Here, $\MPl$ is the reduced Planck mass.
If, however, ALP trapping delays the onset of ALP oscillations, $\Tosc$ can be sufficiently suppressed, leading to
\begin{equation}
    \frac{\tilde{\omega}_\rmii{T}(\Tosc)}{a \Hosc} \sim \frac{\tilde{\omega}_\rmii{T}(\Toscmis)}{a \Hosc}  \rsc^{-2} > 1 \, .
\end{equation}
The amount of supercooling required to open the tachyonic band scales as~\cite{Gerlach:2025fkr}
\begin{equation}\label{eq:rscmin}
    \rsc \propto \left(\frac{\mphi}{\MPl}\right)^\frac{1}{2} \, .
\end{equation}
We note that this relation receives corrections once fermion masses are included in~\cref{eq:dispersion_full} for temperatures below the electron mass, $T < m_e$.
While this modifies the amount of supercooling required, all other results remain unaffected.
We therefore leave a dedicated numerical study of the photon self-energy in the low-temperature limit to future work.\footnote{See \cite{Comadran:2021pkv,Ferreira:2023cqw} for the case of small fermion masses.}

\begin{figure*}
    \centering
    \includegraphics[width=0.9\linewidth]{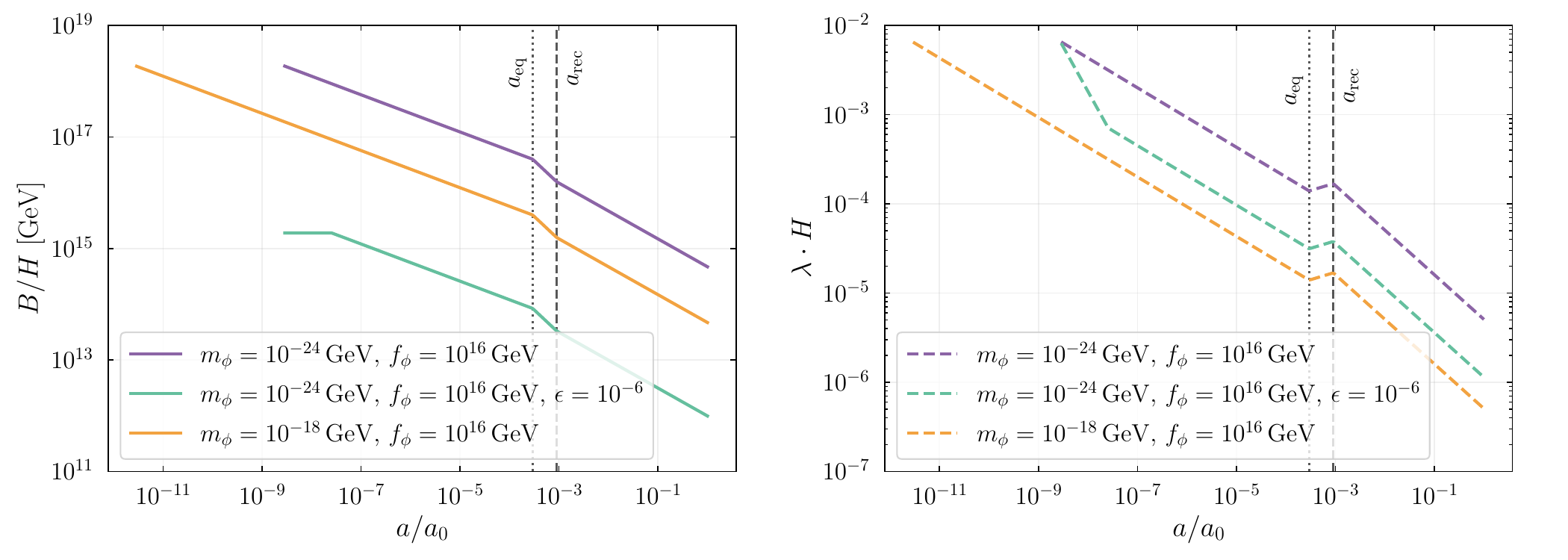}
    \caption{Relative evolution of magnetic field $B$ and (physical) correlation length $\lambda$ compared to initial values at $a_\star$ for three parameter sets, normalized with $H^{-1}$ and $H$, respectively. For each of them, $\alpha=2\alphamin=2$ and $\theta=1$. The origin of the initial values is discussed in \cref{sec:magnetogenesis}.
    The green line starts with an initially     suppressed B-field, which can change the first period of the evolution, since the eddy scale has to catch up first. We set $\epsilon=1$ if not stated otherwise.}
    \label{fig:magnetic_evolution}
\end{figure*}

We then further follow~\cite{Gerlach:2025fkr} and impose a sufficiently small supercooling ratio $\rsc$ for tachyonic growth to become efficient,\footnote{Gauge field production is further damped by the conductivity of the plasma $\sigma \sim T$~\cite{Baym:1997gq,Long:2015cza}, recently also pointed out in~\cite{Sharma:2026hzp}. The amount of supercooling required to suppress the Debye mass of the photon~\eqref{eq:rscmin}~(see Eq.~(3.5) in~\cite{Gerlach:2025fkr} for the full expression) is of the same order as that needed to achieve $\tilde{\omega}_\rmii{T} > \sigma$.} while respecting cosmological bounds from, e.g., big bang nucleosynthesis~(BBN).
In a large part of parameter space, this leads to an intermediate period of thermal inflation, where the false vacuum energy of the trapped ALP drives the expansion rate of the Universe. 
This occurs at the temperature $T_i$, where $\rho_r(T_i)~=~\Delta V~\approx~(\theta \mphi \fphi)^2/2$, with $\rho_r$ being the radiation energy density. Then,
\begin{equation}\label{eq:Ti}
    T_i \simeq \left(\frac{15}{\pi^2 g_{\star,i}}\right)^\frac{1}{4} \left(\theta \mphi \fphi\right)^\frac{1}{2} \, ,
\end{equation}
where $g_{\star,i}$ denotes the relativistic degrees of freedom.

As the ALP eventually starts to oscillate, long-wavelength photon quanta are amplified from the thermal bath.
Tachyonic growth continues until almost the entire energy density of the ALP is transferred to the photon.\footnote{A potential concern is that tachyonic gauge-field production may terminate prematurely due to rapid thermalization of the produced modes. However, such effects are expected to be significantly more relevant for non-Abelian gauge sectors~\cite{Berghaus:2026jkw}, where gauge self-interactions efficiently drive equilibration, than for the Abelian case considered here.}
This produces a helical population of gauge fields that reheat the supercooled Universe to $\Trh = T_i$ and induce large-scale anisotropies in the energy momentum tensor of the primordial Universe, generating a chiral stochastic GW background~\cite{Gerlach:2025fkr}.
In the following, we show that the magnetic fields produced through the tachyonic resonance can survive until today and constitute intergalactic magnetic fields~\cite{Ando:2010rb,Tavecchio:2010mk,doi:10.1126/science.1184192,Essey:2010nd,Chen:2014rsa,Fermi-LAT:2018jdy}, providing a complementary probe of the audible axion model.

\section{Magnetic field evolution}
\label{sec:evolution}

Before studying axion-induced magnetogenesis in our setup, let us first summarize the cosmological evolution of magnetic fields in the presence of a plasma. Two quantities need to be tracked: the initial coherence length $\lambda_\star$ and the field amplitude $B_\star$. As it turns out, for our simple setup, we can already follow the evolution outlined in \cite{Domcke:2025itc}. More details on the evolution are presented in \cite{Brandenburg:1996fc,Brandenburg:1996sa,Olesen:1996ts,Banerjee:2004df,Candelaresi:2011pg,Durrer:2013pga,Brandenburg:2016odr,Brandenburg:2017neh, Kamada:2020bmb}. 
Similar setups have been discussed in \cite{Brandenberger:2025gks,Kamali:2026tgq}. 

The adiabatic evolution in a FLRW universe of the coherence length is just that of any physical scale,
\begin{align}
    \lambda \propto a\,,
\end{align}
and, as a consequence of magnetic flux conservation, $\Phi_B=\int\vect{B}\cdot d\vect{A}=\mathrm{const.}$, the amplitude has to scale with
\begin{align}
    B \propto a^{-2}\,.
\end{align}
This adiabatic evolution is valid in radiation domination~(RD), as long as turbulence effects do not affect the scale of interest. The scale at which turbulences form is denoted eddy scale. It is typically given in terms of the Alfvén velocity \cite{Banerjee:2004df},
\begin{align}
    v_{\rm A}\simeq \frac{B}{\sqrt{\rho+P}}\,,
    \label{eq:Alfven_velo}
\end{align}
and reads 
\begin{align}
    \led\simeq v_{\rm A}\cdot t\simeq \frac{v_{\rm A}}{2H}\,.
    \label{eq:eddy_scale}
\end{align}
As long as the physical coherence length is larger than the eddy scale, the evolution is adiabatic. In RD, however, the eddy scale evolves as $\led \propto a^2$ and thus can catch up with the coherence length. The scale factor at which this occurs can be found by comparing the scales at the time of field generation,
\begin{align}
    \aed = \frac{\lambda_\star}{\led(\astar)}a_\star\,.
\end{align}
If the initial coherence length is already smaller than the eddy scale, the first adiabatic regime is skipped.

What follows is the \textit{cascade regime}, where turbulence effects drive the field quickly towards a maximally helical configuration \cite{Brandenburg:1996fc,Brandenburg:1996sa,Olesen:1996ts}.
From there on, the coherence length follows the eddy scale, which, due to \eqref{eq:Alfven_velo} and \eqref{eq:eddy_scale}, yields 
$\lambda/B\propto a^4$.
Before maximal helicity is reached, the scaling follows \cite{Brandenburg:2016odr,Domcke:2025itc}
\begin{align}
    \lambda\propto a^{2-\frac{\beta+1}{\beta+3}}\,,\quad B\propto a^{-2-\frac{\beta+1}{\beta+3}}\,,
\end{align}
where the value of $\beta$ depends on which mechanism dominates in the inverse cascade (cf.~\cite{Hosking:2020wom,Zhou:2022xhk,Uchida:2022vue,Yanagihara:2023qvx}). 
Since in our scenario the magnetic fields sourced by the tachyonic instability can be assumed to be strongly helical already from creation, the non-helical evolution is not relevant.

Once the maximal helical configuration is reached, the conservation of the total comoving magnetic helicity in the early-Universe plasma will set the scaling. 
We can consider the total magnetic helicity of our field to be $\mathcal{H}=\int d^3x\, \vect{A}\cdot\vect{B}$, and $B=|\nabla\times \vect{A}|$, so $ A \sim \lambda B$. The comoving helicity density then is 
\begin{align}
    h=a^3 \lambda B^2 =\mathrm{const.}
\end{align}
Combined with $\lambda/B\propto a^4$, the scaling for magnetic field strength and correlation length is
\begin{align}
    \lambda\propto a^{5/3}\,,\quad B\propto a^{-7/3}\,.
\end{align}
The increased exponent for $\lambda$ sums up the effect of this \textit{inverse cascade} regime: energy of helical fields is cascaded towards larger scales.
This lasts until recombination, when plasma effects become negligible and adiabatic evolution is recovered.

In summary, the coherence length of an initially strongly helical magnetic field redshifted to today can be written in terms of scale factors
\begin{align}
    \lambda_0=\frac{1}{\arec}\left(\frac{\arec}{\max(\aed,a_\star)}\right)^\frac{5}{3}\max\left(1,\frac{\aed}{a_\star}\right)\lambda_\star\,,
\end{align}
where the \enquote{$\max$} incorporates the fact that the cascade regime can right at emission. The result assumes $\Trh\gg \TBBN$ and $a_0=1$. The corresponding magnetic field can be written as
\begin{align}
    B_0=\arec^{2}\left(\frac{\max(\aed,a_\star)}{\arec}\right)^\frac{7}{3}\left(\max\left(1,\frac{\aed}{a_\star}\right)\right)^{-2}B_\star\,.
\end{align}
An exemplary evolution of an initial magnetic field is shown in~\cref{fig:magnetic_evolution}.

\section{Axion Magnetogenesis}
\label{sec:magnetogenesis}
We now discuss how to obtain the initial coherence length $\lambda_\star$ and field amplitude $B_\star$ at the time of production in terms of the ALP model parameters.
Let us note that tachyonic production of SM photons, as outlined in sec.~\ref{sec:setup}, is only possible in distinct regions of parameter space.
The first requirement is to ensure a radiation-dominated Universe during BBN. 
Therefore, we only include parameter points where $T_i~=~\Trh >~\TBBN$~(cf.~\cref{eq:Ti}).
For simplicity, we impose $\TBBN = 1\,\mathrm{MeV}$.

Next, electromagnetic fields that exceed the Schwinger limit quickly decay into pairs of light SM fermions~\cite{Heisenberg:1936nmg,Schwinger:1951nm,Kobayashi:2014zza,Hayashinaka:2016qqn,Gould:2017fve,Lozanov:2018kpk,Domcke:2018eki,Domcke:2019qmm,Domcke:2021yuz}.
Schwinger pair production sets an upper limit on the energy density that can be transferred to the gauge fields, and hence reduces the efficiency of the tachyonic resonance in a large part of parameter space.
The Schwinger decay rate is suppressed as~\cite{Domcke:2019qmm} 
\begin{equation}
    \Gamma_{\gamma\to f\Bar{f}} \propto \exp\left(-\frac{\pi m_e^2}{eE_\star}\right) \, , 
    \label{eq:Schwinger_rate}
\end{equation}
where $e$ is the electromagnetic charge, $E$ denotes the amplitude of the produced electric field, and $m_e$ is the electron mass.
Given that the ALP temporarily dominates the energy density of the Universe at the time of production in most of the parameter space, and the entire ALP energy density is converted into photons, we can relate $\rho_{\gamma,\star} \sim \rho_{\phi,\mathrm{osc}} = 3 H_\star^2 \MPl^2$, hence
\begin{equation}
    E_\star \sim H_\star \MPl \sim \Trh^2 \, . 
\end{equation}
Consequently, Eq.~\eqref{eq:Schwinger_rate} becomes 
\begin{equation}
    \Gamma_{\gamma\to f\Bar{f}} \propto \exp(-\frac{\pi m_e^2}{e \Trh^2}) \, ,
\end{equation}
which naturally dictates two viable regimes for very small and very large ALP masses, respectively. 
Since $\Trh~\sim~(\mphi\fphi)^\frac{1}{2}$, the scale of photon production is close to BBN for light ALPs. 
Then, $\Trh\sim \TBBN \gtrsim m_e$ and Schwinger production is suppressed.
Very large ALP masses necessitate less supercooling~(cf.~eq.~\eqref{eq:rscmin}), such that $\Tosc \lesssim \Trh$. 
Taking into account the thermal electron mass $m_{e,\rmi{th}} \sim e\Tosc$ then again leads to a suppression of the Schwinger effect.\footnote{Note that the heavy ALP regime still requires a more careful study of the finite-$T$ Schwinger effect.}

To determine the viable parameter space more precisely, we follow refs.~\cite{Domcke:2019qmm,Gerlach:2025fkr}.
For a given axion mass $\mphi$ and decay constant $\fphi$, we determine a minimum axion-photon coupling to ensure tachyonic growth completes. 
To this end, we employ eq.~\eqref{eq:peak_growth_rate_SM} to compute the time when the axion energy density is fully depleted into gauge quanta. 
We require this to occur before the peak wave number is redshifted out of the instability band.
In most of the parameter space, $\alpha_\rmi{min} = 1$~\cite{Gerlach:2025fkr}.
To be more conservative, we set $\alpha = 2\alpha_\rmi{min}$ in the following.
Assuming dynamical equilibrium between the photon, the axion, and the fermions, we compute closed contours of $E$-$B$ pairs that are allowed for a given $(\mphi, \fphi, \alpha)$ in the presence of the Schwinger effect.
We maximize over the contour to obtain an upper bound on the gauge field energy density, which we subsequently employ to determine the regions of parameter space which predict a consistent cosmological evolution; see \cite{Gerlach:2025fkr} for details.

Within the allowed bounds, we may now compute the $B$-field amplitude and coherence length at the time of production.
First, let us note that in principle, it takes a finite amount of time from the onset of ALP oscillations to the moment when the ALP energy density is depleted into gauge quanta.
In the parameter space we consider, this growth time is orders of magnitude smaller than a Hubble time~\cite{Gerlach:2025fkr}, such that one can assume $\rhophistar\simeq \rhophiosc  $.
Therefore, we evaluate all quantities at the oscillation onset.

To obtain the magnetic field amplitude, let us consider the inital ALP energy density
\begin{equation}
    \rhophistar = \frac{(\theta\mphi\fphi)^2}{2} \, .
\end{equation}
This entire energy budget is transferred to the photon with
\begin{equation}
    \rho_{\gamma,\star} = \frac{1}{2} (E_\star^2 + B_\star^2) \, .
\end{equation}
Allowing for different fractions of energy density in form of $E$- and $B$-fields, respectively, we set
\begin{equation}
    B_\star = \sqrt{2 \epsilon \rho_{\gamma,\star}} = \sqrt{\epsilon} \theta \mphi \fphi \, .
    \label{eq:B_star}
\end{equation}
Our numerical study~\cite{Gerlach:2025fkr} shows that the tachyonic resonance typically produces magnetic fields that are significantly stronger than the corresponding electric fields.
Therefore, we set $\epsilon = 1$ in the main part of this article.
In \cref{app:varying_eps}, we show that our results remain robust if $\epsilon$ were to be much smaller.
The corresponding coherence length is dictated by the wave number which experiences the fastest tachyonic growth~\cite{Gerlach:2025fkr},
\begin{equation}
    \tilde{k}_\star =  \theta \alpha m_\phi a_\star \times \begin{cases}\displaystyle
        2/3 \, ,\;\mathrm{if} \; \tilde{\omega}_\rmii{T} < \tilde{k}_\star \, ,\\
        \displaystyle
        1/2\,, \;\mathrm{if} \; \tilde{\omega}_\rmii{T} \geq \tilde{k}_\star \, .
    \end{cases}
\end{equation}
Here, the first case applies at high temperature, where the tachyonic growth rate is suppressed with respect to the zero-temperature scenario~\cite{Hook:2016mqo}.
In the second case, supercooling is sufficiently strong, such that the Debye mass is subdominant compared to fastest growing wave number.
Then Eq.~\eqref{eq:peak_growth_rate_SM} breaks down and both the growth rate and dominant wave number reduce to the vacuum case~\cite{Gerlach:2025fkr}.
In the parameter space we consider in the subsequent section we have $\tilde{k}_\star~\ll~\tilde{\omega}_\rmii{T}$, i.e., the second case applies, such that the magnetic field coherence length reads
\begin{equation}
    \lambda_\star = \frac{2\pi a_\star}{k_\star} = \frac{4\pi}{\theta \alpha \mphi} \, .
\end{equation}

\section{Results}
We now apply the evolution outlined in \cref{sec:evolution} to the initially produced magnetic field from the previous section. 
For $\epsilon\sim\mathcal{O}(1)$, and for the parameter range allowed for the setup, the criterion
\begin{equation}
    \frac{\aed}{\astar} \sim \frac{\lambda_\star}{\lambda_{\rm ed}(\astar)} \sim\frac{4\pi}{\sqrt{12\epsilon}} \frac{\fphi}{\alpha \MPl} \ll 1 \, ,
    \label{eq:scaling_condition}
\end{equation}
is always valid and the initial correlation length is smaller than the eddy scale, meaning the evolution starts with eddy scaling of helical fields. In \cref{app:varying_eps}, we examine the effect of smaller $\epsilon$ more closely.

\begin{figure*}[t]
    \centering
    \includegraphics[width=\linewidth]{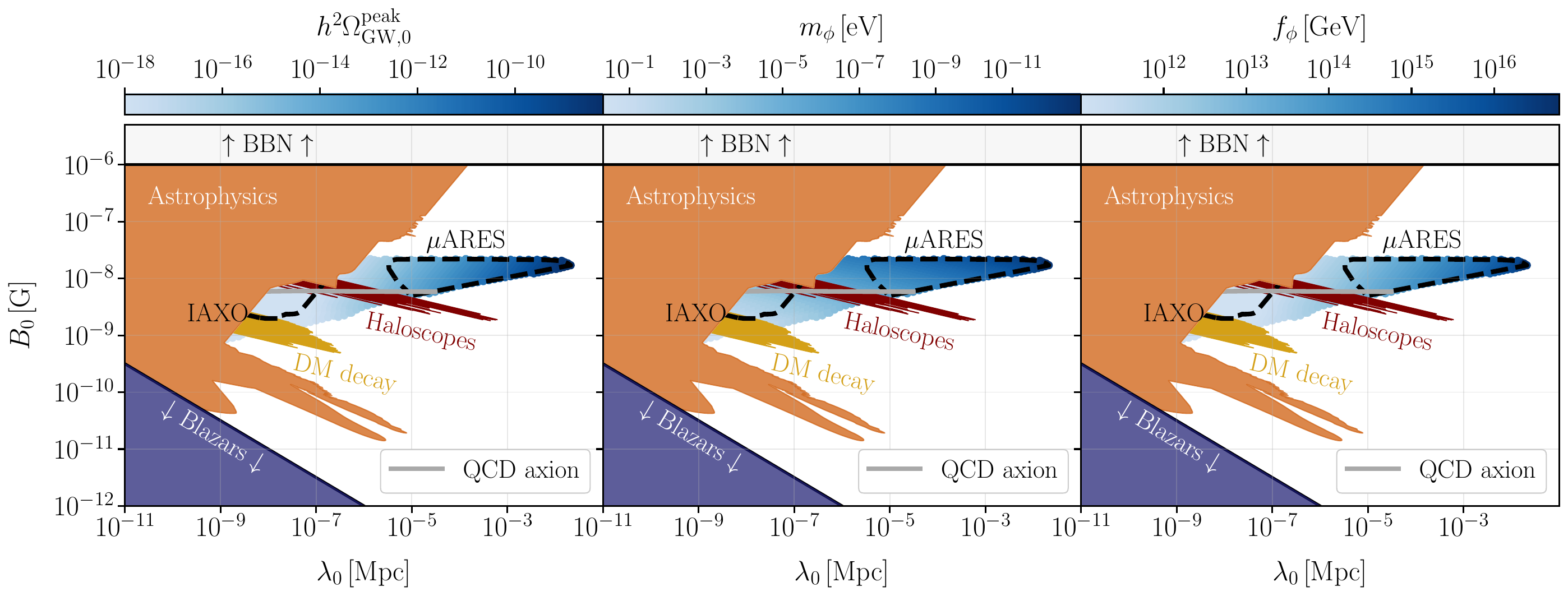}
    \caption{Today's magnetic field strength $B_0$ and coherence length $\lambda_0$. The blue triangles show the parameter space that predicts a consistent evolution of the early Universe~(see~\cite{Gerlach:2025fkr} for details). The color coding indicates the present peak amplitude of the stochastic GW background $h^2\Omega_{\rmii{GW},0}^\rmi{peak}$~(left), the ALP mass~$\mphi$~(center), and the ALP decay constant $\fphi$~(right), respectively. Generally, the produced $B$-fields are sufficiently strong to explain blazar observations~(purple)~\cite{Taylor:2011bn} and well below the upper bound from BBN~\cite{Grasso:2000wj,Kawasaki:2012va}. The remaining colored regions denote current ALP exclusion bounds~\cite{AxionLimits} from astrophysics~\cite{Ayala:2014pea,Dolan:2022kul,Noordhuis:2022ljw,Dessert:2021bkv,Langhoff:2022bij,DeRocco:2022jyq,Dessert:2022yqq,Ruz:2024gkl,Ning:2024eky,Candon:2024eah,Benabou:2025jcv}, dark matter decay into photons~\cite{Grin:2006aw,Regis:2020fhw,Todarello:2023hdk,Wang:2023imi,Janish:2023kvi,Pinetti:2025owq,Saha:2025any}, and haloscopes~\cite{DePanfilis:1987dk,Wuensch:1989sa,ADMX:2018gho,ADMX:2019uok,ADMX:2021nhd,ADMX:2025vom}. The dashed black lines mark the parameter space accessible to the future helioscope IAXO~\cite{IAXO:2019mpb} and GW observatory $\mu\mathrm{ARES}$~\cite{Sesana:2019vho}. To translate the ALP bounds into the $\lambda_0-B_0$ plane, we use $g_\rmi{rh}\sim 10$ in Eqs.~\eqref{eq:lam0_analytic} and \eqref{eq:B0_analytic}, correspdonding to the typical regime of $\Trh$. Parts of the parameter space allow for magnetogenesis induced by the QCD axion~(gray line).
    }
    \label{fig:results}
\end{figure*}

Under this premise, we can apply the evolution analytically to the initial $\lambda_\star$ and $B_\star$, which yields 
\begin{widetext}
\begin{align}
    \displaystyle
    \lambda_0 = \lambda_\star \left(\frac{a_\rmi{rec}}{a_\rmi{rh}}\right)^\frac{5}{3} \frac{a_0}{a_\rmi{rec}} 
        &= 0.50 \times 10^{30} g_\rmi{rh}^{\frac{5}{36}} \left(\alpha \theta^\frac{1}{6}\right)^{-1} \left(\frac{\fphi}{10^{10}\,\GeV}\right)^\frac{5}{6} \left(\frac{\mathrm{eV}}{\mphi}\right)^\frac{1}{6}\,\GeV^{-1} \nonumber \\ \label{eq:lam0_analytic}
        &= 3.20 \times 10^{-9} g_\rmi{rh}^{\frac{5}{36}} \left(\alpha \theta^\frac{1}{6}\right)^{-1} \left(\frac{\fphi}{10^{10}\,\GeV}\right)^\frac{5}{6} \left(\frac{\mathrm{eV}}{\mphi}\right)^\frac{1}{6}\,\mathrm{Mpc} \, , \\[0.5em]
    \displaystyle 
    B_0 = B_\rmi{rh}  \left(\frac{a_\rmi{rec}}{a_\rmi{rh}}\right)^{-\frac{7}{3}} \left(\frac{a_0}{a_\rmi{rec}}\right)^{-2} 
        &= 5.44 \times 10^{-29} g_\rmi{rh}^{-\frac{7}{36}} \theta^{-\frac{1}{6}} \left(\frac{10^{10}\,\GeV}{\fphi}\right)^\frac{1}{6} \left(\frac{\mathrm{eV}}{\mphi}\right)^\frac{1}{6} \GeV^2 \nonumber\\\label{eq:B0_analytic}
        &= 27.84 \times 10^{-10} g_\rmi{rh}^{-\frac{7}{36}} \theta^{-\frac{1}{6}} \left(\frac{10^{10}\,\GeV}{\fphi}\right)^\frac{1}{6} \left(\frac{\mathrm{eV}}{\mphi}\right)^\frac{1}{6} \mathrm{G}\,.
\end{align}
\end{widetext}
From this expression, it becomes obvious why it suffices to consider small $\mphi$.
In~\cite{Gerlach:2025fkr}, we found that ultra-heavy ALPs with $\mphi~\sim~\mathcal{O}(10^{10})\,\GeV$ allow for tachyonic photon production, producing a sizable GW background in the $\mathrm{GHz}$ regime.
Due to the extended eddy scaling, the relic magnetic field strength decreases to $B_0\sim\mathcal{O}(10^{-14})\,\mathrm{G}$, i.e., is too weak to explain blazar observations~(see below).

Therefore, we only consider the small-$m_\phi$ region identified in~\cite{Gerlach:2025fkr}, which approximately corresponds to model parameters in the range
\begin{align}
1.5 \times 10^{-13}\,\mathrm{eV} &\lesssim \mphi \lesssim 10^{-3}\,\mathrm{eV} \, ,\\[0.5em]
10^{12}\,\mathrm{GeV} &\lesssim \fphi \lesssim 6 \times 10^{16}\,\mathrm{GeV} \, .
\end{align}
For each parameter set $(\mphi,\fphi,2\alphamin)$ we evolve the $B$-field amplitude and coherence length to today, and plot the endpoints of the evolution, which form the blue triangles in the $\lambda_0-B_0$ plane in \cref{fig:results}.
Interestingly, all resulting points are well below the upper limit from BBN~\cite{Grasso:2000wj,Kawasaki:2012va}, while still yielding sufficiently large amplitudes to account for intergalactic magnetic fields inferred from blazar observations~\cite{Ando:2010rb,Tavecchio:2010mk,doi:10.1126/science.1184192,Taylor:2011bn,Essey:2010nd,Chen:2014rsa,Fermi-LAT:2018jdy}.

It is worth noting that in the entire parameter space, magnetic field production occurs at temperatures well below the electroweak crossover, $\Trh~\sim~1\,\mathrm{MeV}~-~20\,\mathrm{GeV}$.
This avoids the strong bounds from baryon asymmetry overproduction by helical hypermagnetic fields \cite{Kamada:2016eeb,Kamada:2016cnb}.

The blue color coding indicates GW amplitude $h^2\Omega_{\rmii{GW},0}^\rmi{peak}$~(left), ALP mass $\mphi$~(center), and decay constant $\fphi$~(right), respectively.
The computation of the GW signal is outlined in~\cite{Gerlach:2025fkr}.
Note that the largest correlation lengths of $\lambda_0 \sim \mathcal{O}(10^{-2})\,\mathrm{Mpc}$ are related to small $\mphi$ and large $\fphi$.
This also marks the regime with the most promising GW detection prospects, indicated by the projected sensitivity region of the future observatory $\mu$ARES~\cite{Sesana:2019vho}~(black dashed).
In addition, we display several existing ALP constraints as colored areas, corresponding to limits from astrophysics~\cite{Ayala:2014pea,Dolan:2022kul,Noordhuis:2022ljw,Dessert:2021bkv,Dessert:2022yqq,Benabou:2025jcv,Ruz:2024gkl}~(orange), dark matter decay into photons~\cite{Grin:2006aw,Regis:2020fhw,Todarello:2023hdk,Wang:2023imi,Janish:2023kvi,Pinetti:2025owq,Saha:2025any}~(yellow), and haloscopes~\cite{DePanfilis:1987dk,Wuensch:1989sa,ADMX:2018gho,ADMX:2019uok,ADMX:2021nhd,ADMX:2025vom}~(red).
Those are obtained by recasting the limits~\cite{AxionLimits} from the $m_\phi - g_{\phi\gamma\gamma}$ plane, assuming a KSVZ ALP~\cite{Kim:1979if,Shifman:1979if} and taking $\alpha = 2\alphamin$. 
In the same way, we obtain the projected sensitivity of the future helioscope IAXO~\cite{IAXO:2019mpb}~(black dashed).
Lastly, we indicate the QCD axion by the gray solid lines, which is, however, largely constrained.
The remaining parameter space, on the other hand, is almost entirely falsifiable by future helioscopes and the (non-)observation of $\mu$Hz GWs with $\mu$ARES.

Let us finally note that our model is further constrained by the requirement to produce the correct relic dark matter abundance.
In the relevant parameter space, this requires a relative suppression of the ALP energy density through the tachyonic resonance by a factor of $\mathcal{O}(10^{-10}-10^{-6})$~\cite{Gerlach:2025fkr}.
Previous lattice studies~\cite{Ratzinger:2020oct} find a suppression of $\mathcal{O}(10^{-2})$ for $\alpha \sim 50-100$. 
In the absence of further lattice computations, it is currently unclear whether our employed axion-photon coupling of $\alpha \sim 1-2$ allows for a greater suppression of the ALP abundance.
If that is not the case, further model building, such as a time-varying ALP mass, can yield the correct relic abundance~\cite{McAllister:2008hb,Silverstein:2008sg,Hebecker:2014eua,McAllister:2014mpa,Blumenhagen:2014gta,Marchesano:2014mla}.

\section{Conclusions}
In this work, we show that ALPs coupled to the SM photon provide a framework for cosmic magnetogenesis, realized through tachyonic photon production in the trapped misalignment mechanism.
Taking into account all relevant astrophysical and cosmological bounds on our scenario, we find that ALPs with masses $1.5~\times~10^{-13}\,\mathrm{eV}~\lesssim~\mphi~\lesssim 10^{-3}\,\mathrm{eV}$, and decay constants $10^{12}\,\mathrm{GeV} \lesssim \fphi \lesssim 6 \times 10^{16}\,\mathrm{GeV}$, can source intergalactic magnetic fields that exceed the lower bounds from recent blazar observations~\cite{Ando:2010rb,Tavecchio:2010mk,doi:10.1126/science.1184192,Essey:2010nd,Chen:2014rsa,Fermi-LAT:2018jdy}; see Fig.~\ref{fig:results}.

The production of magnetic fields is accompanied by the generation of a stochastic GW background~\cite{Gerlach:2025fkr}.
Intriguingly, the parameter space least constrained by astrophysical ALP bounds predicts GWs in the $\mu\mathrm{Hz}$ regime, rendering it accessible to the future interferometer $\mu$ARES~\cite{Sesana:2019vho}.
In addition, parts of the parameter space will be tested via the axion helioscope IAXO~\cite{IAXO:2019mpb}.
This highlights the compelling multi-messenger discovery potential of the audible axion model.

\begin{figure*}
    \centering
    \includegraphics[width=0.6\linewidth]{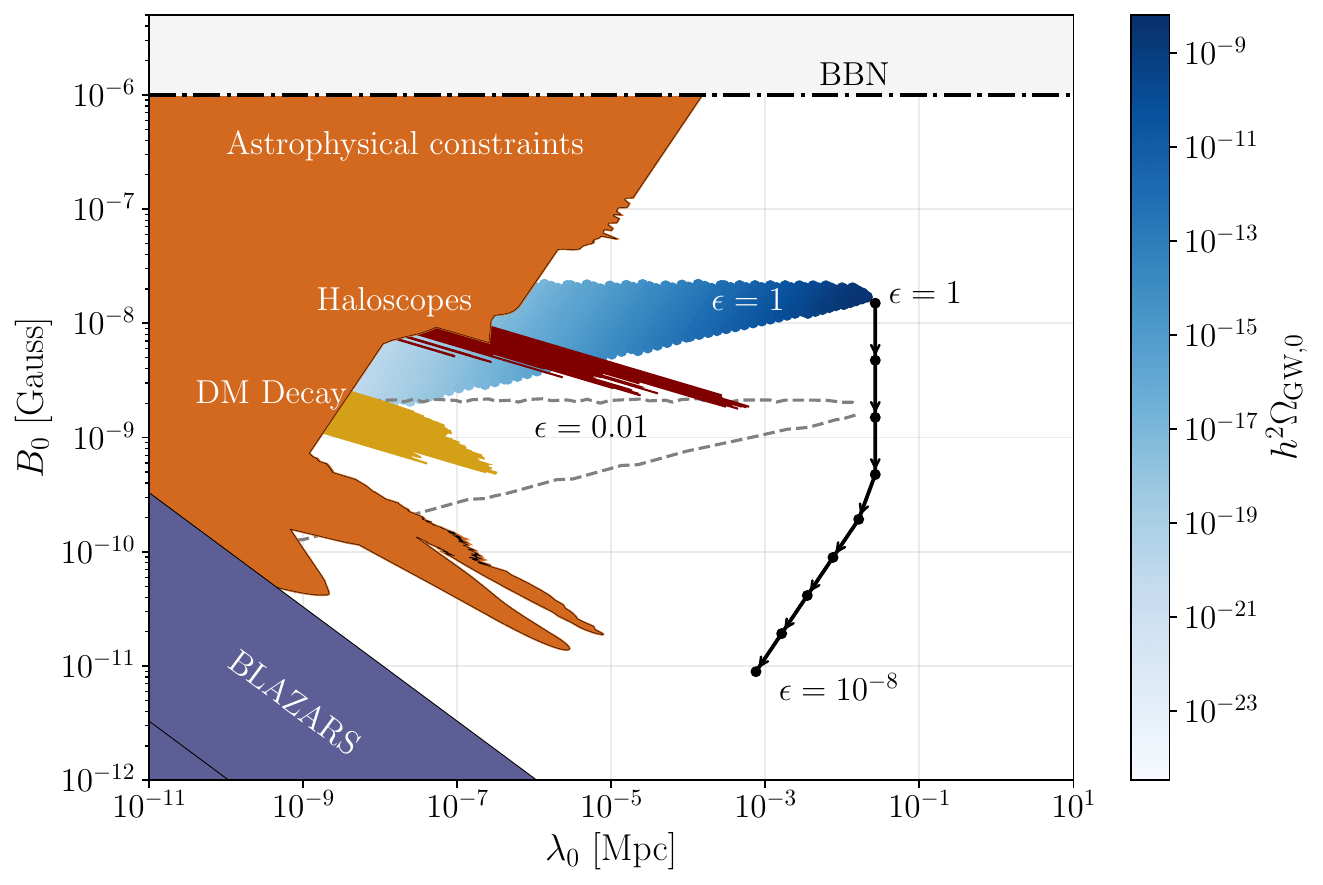}
    \caption{
    The effects of small $\epsilon$: If only a fraction of the available energy density sources the magnetic field $B_\star$, the redshifted $B_0$ scales with $\sqrt{\epsilon}$. We show the allowed parameter space for $\epsilon=1$ and $\epsilon=0.01$ (gray envelope). Only for $\fphi$ closer to $\MPl$, a small $\epsilon$ also changes the initial scaling of the physical correlation length. Only for the largest parameter points, combined with very small $\epsilon$ below $10^{-3}$, the correlation length is affected by an era of adiabatic scaling before the eddy scale takes over. This changed evolution affects the endpoint in the $\lambda_0-B_0$ plane. Moving to smaller $\epsilon$, the right tip of the triangle is crunched at the black line, which shows one parameter point evaluated for different $\epsilon$, while everything left of it is just affected in $B_0$. We show the same axion constraints as in \cref{fig:results}, corresponding to limits from astrophysics (orange), haloscopes (red) and DM decay (yellow).}
    \label{fig:epsilon}
\end{figure*}

To obtain a complete and self-consistent framework, future work should incorporate a viable mechanism for baryogenesis. 
Several avenues merit exploration. 
First, reducing the required amount of supercooling could allow a pre-existing baryon asymmetry to persist. 
To this end, the effect of fermion masses below $T < m_e$ on the photon dispersion relation~\eqref{eq:dispersion_full} should be taken into account.
In contrast, if supercooling remains substantial, any initial asymmetry would be erased by entropy injection during reheating. 
Then, one may explore baryogenesis driven by axion dynamics, for instance within axiogenesis-inspired scenarios~\cite{Co:2019wyp,Co:2022kul}. 
We leave a detailed study of these possibilities to future work.

\begin{acknowledgments}
We thank M.~Lewicki and D.~Perri for drawing our attention to intergalactic magnetic fields, and P.~Schicho and K.~Seppänen for correspondence on the photon dispersion relation.
We also thank V. Domcke for pointing out constraints from baryon asymmetry overproduction. 
PS and CG acknowledge support by the Cluster of Excellence “Precision Physics, Fundamental Interactions, and Structure of Matter” (\textit{PRISMA}$^{++}$ EXC 2118/2) funded by the Deutsche Forschungsgemeinschaft (DFG, German Research Foundation), (Project No. 390831469).
DS is funded by the DFG through the Emmy Noether Programme Project No. 548044346. 
DS also acknowledges support by the DFG through the CRC-
TR 211 ’Strong-interaction matter under extreme conditions’– project number 315477589 – TRR 211.
\end{acknowledgments}

\appendix

\section{Small $\epsilon$}\label{app:varying_eps}
Let us briefly consider the case that instead of the magnetic field being much stronger, the electric field is mainly populated, such that $E^2\gg B^2$. In this case, only a small fraction $\epsilon$ of the total energy density available at the time of the instability sources the magnetic fields, cf. Eq.~\eqref{eq:B_star}.

As long as the correlation length at magnetogenesis is well below the eddy scale, this only affects the amplitude of the magnetic field, but not the scaling and therefore neither the cosmological evolution. The resulting $B_0$ is multiplied by $\sqrt{\epsilon}$. In \cref{fig:epsilon}, we show the parameter space from \cref{fig:results} for $\epsilon=1$ (colored) and $\epsilon=0.01$ (gray envelope). Even much lower fractions of magnetic energy density fulfill the blazar bounds in most of the parameter space allowed by the mechanism.

For $\fphi$ closer to the Planck scale and small $\epsilon$ however, the correlation length might start above the eddy scale, rendering an initial era of adiabatic evolution as explained in \cref{sec:evolution}. This changes the resulting $B_0$ \textit{and} $\lambda_0$. In the parameter space allowed for our mechanism, this is not the case for realistic values of $\epsilon$. We show a trajectory (black) for one allowed parameter point at the tip in \cref{fig:epsilon}, where we fix $\mphi$ and $\fphi$ and reduce $\epsilon$ step by step. Once the correlation length is affected, the parameter space on the right is contracted to the black line.

\bibliography{apssamp}

\end{document}